\documentclass[a4paper]{article}

\usepackage{INTERSPEECH2020}
\usepackage{amsmath,amssymb,fixmath}
\usepackage{subfig}
\usepackage{graphicx}
\usepackage{float}
\usepackage{cite}
\usepackage[dvips]{epsfig}
\usepackage{multirow}
\usepackage{makecell}
\usepackage{multicol}
\usepackage{tabularx}
\usepackage{color}

\PassOptionsToPackage{hyphens}{url}\usepackage{hyperref}
\newcommand{\eunwooedit}[1]{\textcolor{black}{#1}}

\newcommand{\minjaeedit}[1]{\textcolor{black}{#1}}
\newcommand{\reunwooedit}[1]{\textcolor{black}{#1}}
\newcommand{\rryuichiedit}[1]{\textcolor{black}{#1}}
\newcommand{\rminjaeedit}[1]{\textcolor{black}{#1}}

\newcommand{\ttsmos}{$4.57$ }

\title{Neural Text-to-Speech with a Modeling-by-Generation Excitation \eunwooedit{Vocoder}}
\name{Eunwoo Song$^1$, Min-Jae Hwang$^2$, Ryuichi Yamamoto$^3$, Jin-Seob Kim$^1$, \\ Ohsung Kwon$^1$, and Jae-Min Kim$^1$}
\address{
  $^1$NAVER Corp., Seongnam, Korea\\
  $^2$Search Solutions Inc., Seongnam, Korea \\
  $^3$LINE Corp., Tokyo, Japan}
\email{eunwoo.song@navercorp.com}

\begin{document}

\maketitle
\begin{abstract}
  
    This paper proposes a modeling-by-generation (MbG) excitation \eunwooedit{vocoder} for a neural text-to-speech (TTS) system. 
    % model structure--> vocoder
    Recently proposed neural excitation vocoders can realize qualified waveform generation by combining a vocal tract filter with a WaveNet-based glottal excitation generator. 
    However, when these vocoders are used in a TTS system, the quality of synthesized speech is often degraded owing to a mismatch between training and synthesis \reunwooedit{steps}. 
    % phases --> steps
    Specifically, the vocoder is separately trained from an acoustic model front-end\rminjaeedit{.} 
    \rminjaeedit{Therefore,} estimation errors of the acoustic model are inevitably boosted throughout the synthesis process of the vocoder back-end. 
    % and --> even though
    % even though -->therefore
    % MJ: Too long sentence --> Separate by 'Therefore'
    To address this problem, we propose to incorporate an MbG structure into the vocoder's training process. 
    In the proposed method, the excitation signal is extracted by the \eunwooedit{acoustic model's} generated spectral parameters, and the neural vocoder is then optimized not only to learn the target excitation's distribution but also to compensate for the estimation errors occurring from the acoustic model. 
    Furthermore, as the generated spectral parameters are shared in the training and synthesis \reunwooedit{steps}, their mismatch conditions can be reduced effectively. 
    % steps --> phases
    % phases --> steps
    The experimental results verify that the proposed system provides high-quality synthetic speech by achieving a mean opinion score of \ttsmos within the TTS framework.

\end{abstract}
\noindent\textbf{Index Terms}: neural text-to-speech, WaveNet, ExcitNet, modeling-by-generation vocoder
\section{Introduction}
\vspace*{5pt}	se
	Generative models for raw speech waveform have significantly improved the quality of neural text-to-speech (TTS) systems \cite{ze2013statistical, Oord2016WaveNetAG}.
	%Specifically, by conditioning acoustic features to the network input, neural vocoding models such as WaveNet, WaveRNN, and WaveGlow have successfully replaced traditional parametric vocoders \cite{Oord2016WaveNetAG, tamamori2017speaker, kalchbrenner2018efficient, prenger2019waveglow}.
	Specifically, by conditioning acoustic features to the network input, neural vocoding models such as WaveNet, WaveRNN, and WaveGlow \eunwooedit{successfully generate a time-sequence of speech signal} \cite{Oord2016WaveNetAG, tamamori2017speaker, kalchbrenner2018efficient, prenger2019waveglow}.
	More recently, neural excitation vocoders such as GlotNet, ExcitNet, \reunwooedit{LP-WaveNet} and LPCNet \cite{juvela2018speaker, song2019excitnet, hwang2018lp, valin2019lpcnet, hwang2020improving} have exploited the advantages of linear prediction (LP)-based parametric vocoders.
    % LP-WaveNet, iLPCNet are added
	In this type of vocoder, an adaptive predictor is used to decouple the formant-related spectral structure from the input speech signal, and the probability distribution of its residual signal (i.e. the excitation signal) is then modeled by the vocoding network. 
	As variation in the excitation signal is only constrained by vocal cord movement, the training and generation processes become much more efficient.
	
	However, because the vocoding and acoustic models have been trained separately, it is not known whether or not combining them within the TTS framework would benefit synthesis quality. 
	Furthermore, as parameters estimated from the acoustic model are used as a direct input of the vocoding model \eunwooedit{in the synthesis step}, estimation errors of the acoustic features can be propagated throughout the synthesis process. 
	It is therefore crucial to model the interactions between the acoustic and vocoding elements during the training process in order to achieve the best complete performance of the TTS system.
	
	In this paper, we propose a neural excitation model based on modeling-by-generation (MbG) in which the \eunwooedit{spectral} parameters generated from the acoustic model are utilized in the neural vocoder's training process.
	Specifically, the target excitation is defined as a combination of the prediction errors from the LP analysis and those from the acoustic model. 
	The vocoding model is then optimized to learn the distribution of the target excitation while compensating for the errors from the acoustic model.
	It has been reported elsewhere that training the neural vocoder with generated acoustic parameters improves synthetic quality \cite{Shen2018NaturalTS}.
	Although the MbG method is similar to this approach, there are also clear differences in that MbG aligns \eunwooedit{even} the target excitation signal with the acoustic model's generated spectral parameters.
	
	We investigated the effectiveness of the proposed method by conducting subjective evaluation tasks. 
	The MbG structure can be extended to any neural excitation vocoder that uses LP coefficients, but the focus here is on the WaveNet-based ExcitNet vocoder \cite{song2019excitnet}.
    The experimental results show that a TTS system with the proposed MbG-ExcitNet vocoder provides significantly better perceptual quality than a similarly configured system with a conventional vocoder. In particular, our TTS framework achieves \ttsmos mean opinion score (MOS).
    
%%%%%%%%%%%%%%%%%%% Fig: ExcitNet %%%%%%%%%%%%%%%%%%%%%%%%%%%%%%
	\begin{figure*}[!t]
	\begin{minipage}[t]{.33\linewidth}
	\centerline{\epsfig{figure=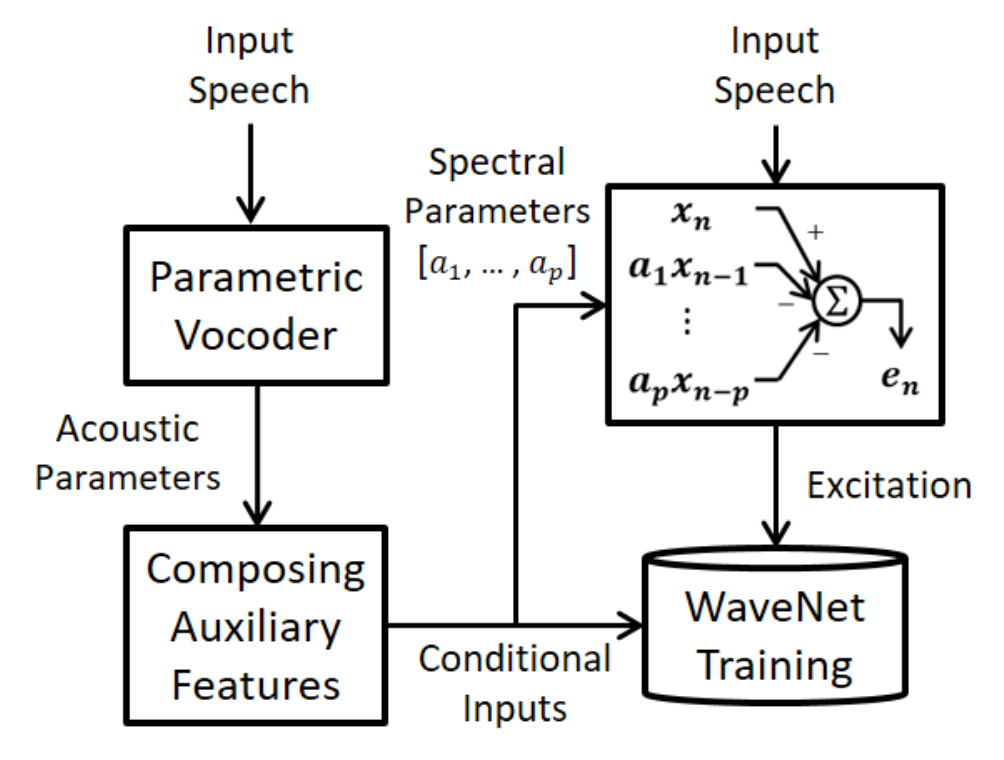,width=53mm}}
	%\vspace*{-1pt}	
	\centerline{(a)}  \medskip
	\end{minipage} 
	\begin{minipage}[t]{.33\linewidth}
	\centerline{\epsfig{figure=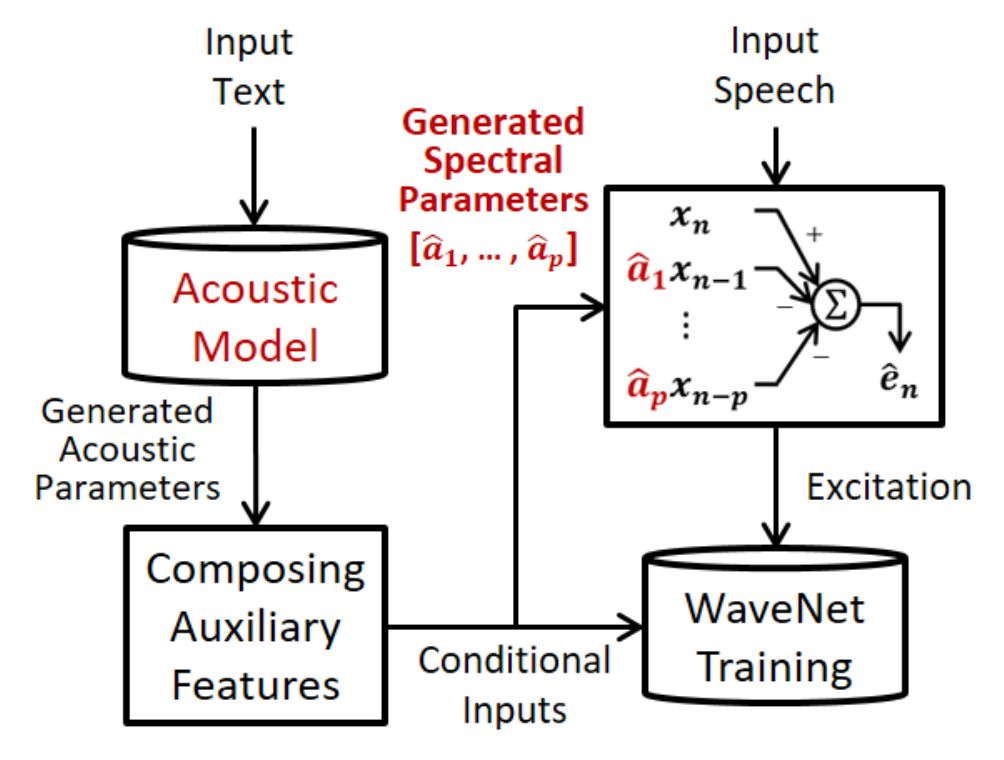,width=53mm}}
	%\vspace*{-1pt}	
	\centerline{(b)}  \medskip
	\end{minipage}	
	\begin{minipage}[t]{.33\linewidth}
	\centerline{\epsfig{figure=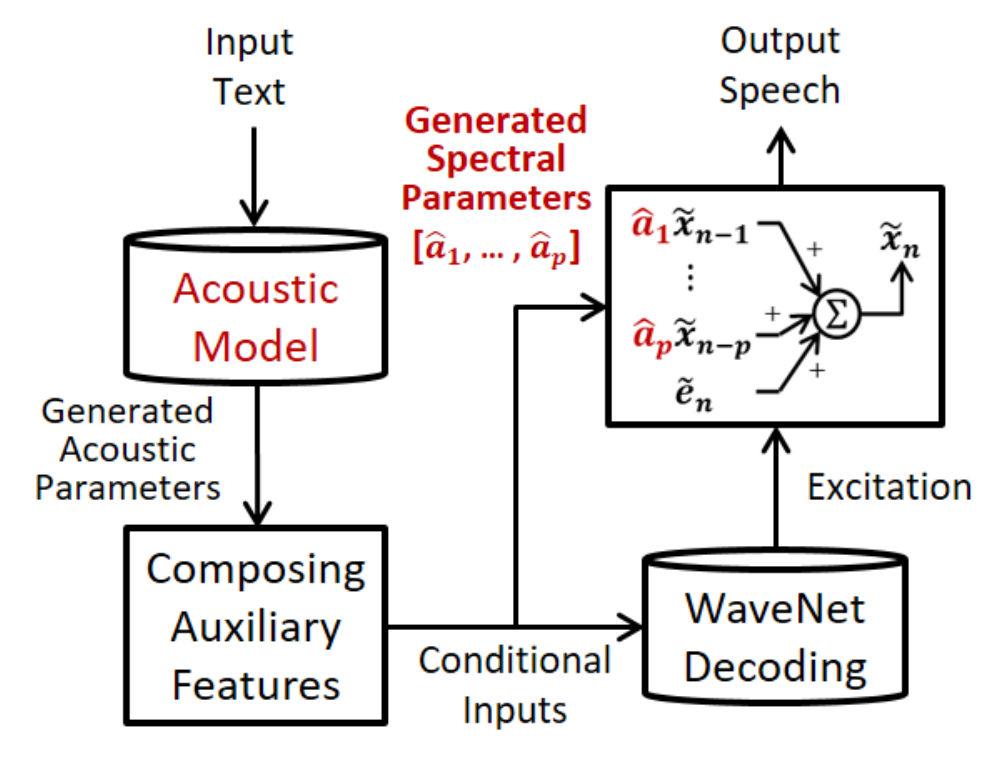,width=53mm}}
	%\vspace*{-1pt}	
	\centerline{(c)}  \medskip
	\end{minipage}	
	%\vspace*{-6pt} 
	\caption{An ExcitNet vocoder for a TTS system: (a) conventional training; (b) proposed MbG training; and (c) synthesis methods.}	
	%\vspace*{7pt} 
	\label{fig:ExcitNet}
	\end{figure*}
%%%%%%%%%%%%%%%%%%% Fig: ExcitNet %%%%%%%%%%%%%%%%%%%%%%%%%%%%%%

\section{Related work}
\label{sec:related work}
\vspace*{5pt}

    The idea of using an MbG structure is not new. 
    In a study of parametric \textit{glottal vocoders}, Juvela et al. \cite{juvela2017reducing} first proposed the closed-loop extraction of glottal excitation from the generated spectral parameters, and our own previous work proposed the MbG structure to compensate for missing noise components in generated glottal signals \cite{hwang2018modeling}. 
    However, it was not possible to fully utilize the effectiveness of the MbG training strategy because our experiments were only performed with simple deep learning models including stacked feed-forward and/or long short-term memory (LSTM) networks.
    
    Our aim here was to extend \minjaeedit{the usage} of the MbG structure to recently proposed neural excitation models (e.g. ExcitNet) with autoregressive acoustic models (e.g. Tacotron) \cite{wangIS2017tacotron,  Shen2018NaturalTS, okamoto2019tacotron}. 
    % use --> the usage
    As the accuracy of acoustic models has been significantly improved, it is now possible to extract stable excitation signals from the generated spectral parameters. 
    Furthermore, the ExcitNet vocoder directly models the time-domain excitation sequence which enables straightforward application of the MbG structure to the training process. 
    As a result, the entire model can be stably and easily trained while the perceptual quality of the synthesized speech is significantly improved.

%%%%%%%%%%%%%%%%%%% Fig: nll %%%%%%%%%%%%%%%%%%%%%%%%%%%%%%  
    \begin{figure}[!t]
    \centerline{\epsfig{figure=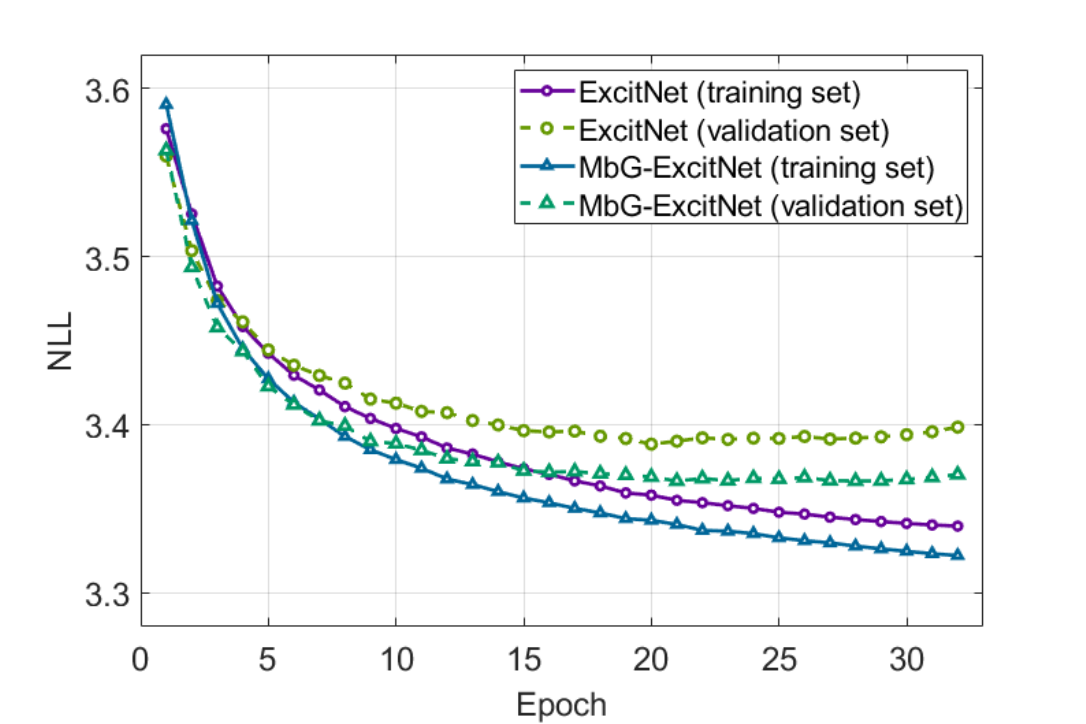,width=72mm}}
%    \vspace*{-6pt}  
    \caption{Negative log-likelihood (NLL) obtained during the training process with respect to the plain ExcitNet and MbG-based ExcitNet (MbG-ExcitNet) training methods.}
%    \vspace*{-10pt}  
    \label{fig:nll}
    \end{figure}
%%%%%%%%%%%%%%%%%%% Fig: nll %%%%%%%%%%%%%%%%%%%%%%%%%%%%%%       

\section{ExcitNet TTS systems}

\subsection{ExcitNet vocoders}
\label{ssec:ch1-1}

    The basic WaveNet framework is an autoregressive network which generates a probability distribution of discrete speech symbols from a fixed number of past samples \cite{van2016wavenet}. 
    % 0729 Ryuichi: removed `merits` as it is duplicate with `advantanges`.
    %The ExcitNet vocoder is an advanced version of this network which takes advantages of both the LP vocoder's and the WaveNet structure's merits.
    % 0729 Eunwoo: vocoder's --> vocoder
    The ExcitNet vocoder is an advanced version of this network which takes advantages of both the LP \reunwooedit{vocoder} and the \rryuichiedit{WaveNet structure}.
    In an ExcitNet framework, an LP-based adaptive predictor is used to decouple the spectral formant structure from the input speech signal (Fig.~\ref{fig:ExcitNet}a).
    The WaveNet model is then used to train the distribution of the prediction residuals (i.e. excitation) as follows:
    \begin{equation}\label{eq:en}
	p(\mathbf{e}|\mathbf{h})=\prod\limits_{n=1}^{N}{p({{e}_{n}}|{{e}_{1}},...,{{e}_{n-1}}, \mathbf{h})},
	\end{equation}
	\begin{equation}\label{eq:lp}
	{e}_{n} = {x}_{n} - \sum_{k=1}^{p}\alpha_{k}x_{n-k},
	\end{equation}
	where $x_n$ and $e_n$ denote the $n^{th}$ sample of speech and excitation, respectively; $\alpha_k$ denotes the $k^{th}$ LP coefficient with the order $p$; $\mathbf{h}$ denotes the conditional inputs composed of acoustic parameters.
    
    In the speech synthesis step (Fig.~\ref{fig:ExcitNet}c), the acoustic parameters of the given input text are generated by \eunwooedit{a pre-trained} acoustic model. 
    These parameters are then used as conditional inputs for the WaveNet model to generate the corresponding time sequence of the excitation signal. 
    Finally, the speech signal is reconstructed by passing the generated excitation signal through the LP synthesis filter.

\subsection{MbG-structured ExcitNet vocoders}
\label{ssec:ch3-2}

    To further improve the quality of the synthesized speech, we propose the incorporation of an MbG structure into the training process of the ExcitNet vocoder. 
    As illustrated in Fig.~\ref{fig:ExcitNet}a, conventional vocoding models are trained separately from the acoustic model, even though the generated acoustic parameters, which contain estimation errors, are used as direct conditional inputs (Fig.~\ref{fig:ExcitNet}c). This inevitably causes quality degradation of the synthesized speech as the estimation errors from the acoustic model are boosted non-linearly throughout the synthesis process in the vocoder back-end.
    
    Fig.~\ref{fig:ExcitNet}b shows the proposed MbG training method which uses closed-loop extraction\footnote{This extraction method has been adopted in analysis-by-synthesis speech coding frameworks \cite{atal1979predictive, drugman2014glottal} where the encoder and decoder share the same quantized filter parameters for minimizing their mismatch conditions.} of the excitation signal. 
    To minimize the mismatch between the training and the generation processes, the LP coefficients in the training \reunwooedit{step} are replaced with those generated by the pre-trained acoustic model as follows:
    % phase --> step
    \begin{equation}\label{eq:lp_gen}
	\hat{e}_{n} = {x}_{n} - \sum_{k=1}^{p}\hat{\alpha}_{k}x_{n-k},
	\end{equation}    
	where $\{{\hat{\alpha}_{1}},...,{\hat{\alpha}_{p}}\}$ denotes the generated LP coefficients.
	By combining equations~(\ref{eq:lp}) and (\ref{eq:lp_gen}), the excitation sequence can be represented as follows:
	\begin{equation}\label{eq:lp_re}
	\hat{e}_{n} = {e}_{n} + {e}^{am}_{n},
	\end{equation}    
	where $e^{am}_n$ denotes an \textit{intermediate prediction} defined as follows:
	\begin{equation}\label{eq:lp_final}
	{e}^{am}_{n} = \sum_{k=1}^{p}(\alpha_{k}-\hat{\alpha}_{k})x_{n-k}.
	\end{equation}
    Using the excitation signal (i.e. $\hat{e}_{n}$) as the training target means that it becomes possible to guide the model to learn the distributions of the true excitation signal (i.e. $e_n$) as well as compensate for the acoustic model's estimation errors (i.e. $e^{am}_n$).
    Furthermore, because the training and synthesis processes share the same LP coefficients, it is also possible to minimize any mismatch.
    
%%%%%%%%%%%%%%%%%%% Fig: am %%%%%%%%%%%%%%%%%%%%%%%%%%%%%%  
    \begin{figure*}[!t]
    \centerline{\epsfig{figure=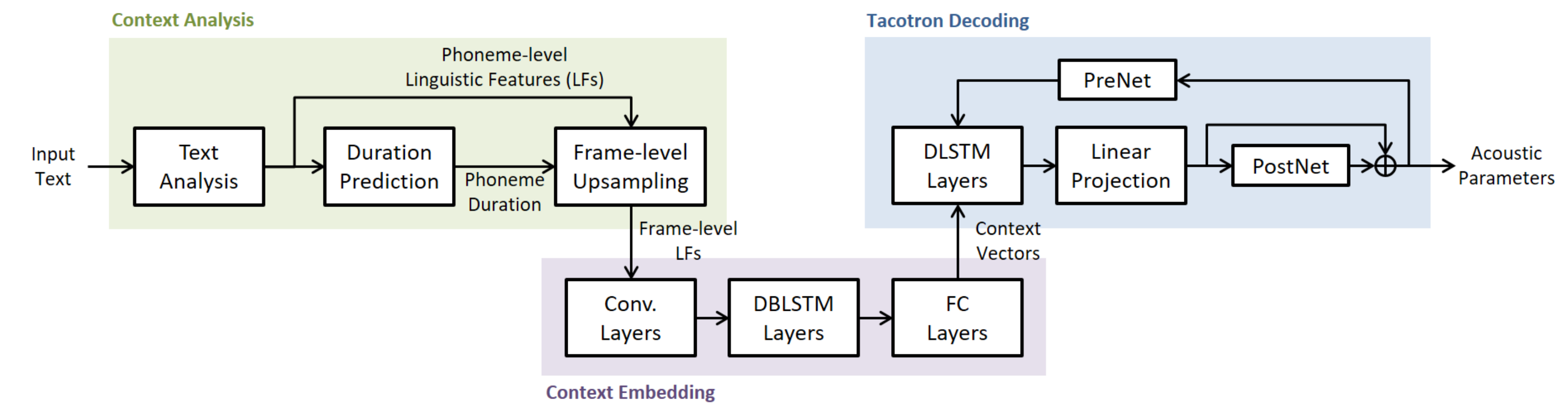,width=164mm}}
%    \vspace*{-6pt}  
    \caption{Acoustic model consisting of three sub-modules: context analysis, context embedding, and Tacotron decoding.}
    %\vspace*{-10pt}  
    \label{fig:am}
    \end{figure*}
%%%%%%%%%%%%%%%%%%% Fig: am %%%%%%%%%%%%%%%%%%%%%%%%%%%%%%    

    The merits of the proposed method are presented in Fig.~\ref{fig:nll} which shows the negative log-likelihood obtained from the training and validation sets. 
    The proposed MbG-ExcitNet model enables a reduction in both training and validation errors as compared to a plain ExcitNet approach. 
    It is therefore expected that the proposed method will provide more accurate training and generation results, to be further discussed in the following section.
    
\section{Experiments}
	\label{sec:experiment}
	\subsection{Experimental setup}
	\subsubsection{Database}
	The experiments used a phonetically and prosodically balanced speech corpus recorded by a Korean female professional speaker. 
	The speech signals were sampled at 24 kHz with 16 bit quantization.
	In total, 4,408 utterances (7.9 hours) were used for training, 230 utterances (0.4 hours) were used for validation, and a further 120 utterances (0.2 hours) were used for testing. 
	The acoustic features were extracted using the improved time-frequency trajectory excitation vocoder at analysis intervals of 5 ms \cite{song2017effective}, and these features included 40-dimensional line spectral frequencies (LSFs), fundamental frequency (F0), energy, voicing flag (v/uv), 32-dimensional slowly evolving waveform (SEW), and 4-dimensional rapidly evolving waveform (REW), all of which constituted a 79-dimensional feature vector.
	
	\subsubsection{Acoustic model}
	Although there are many state-of-the-art acoustic models available, including Tacotron and Transformer \cite{wangIS2017tacotron, Shen2018NaturalTS, li2019close}, we opted to pursue a Tacotron model with phoneme alignment approach \cite{okamoto2019tacotron} because of its fast and stable generation and competitive synthetic quality.
	Fig.~\ref{fig:am} is a block diagram of the acoustic model which consists of three sub-modules, namely context analysis, context embedding, and Tacotron decoding. 
	
	In the context analysis module, the phoneme-level linguistic feature vectors were extracted from the input text.
	These were composed of 330 binary features for categorical linguistic contexts and 24 features for numerical linguistic contexts. 
	Having input these features, the corresponding phoneme duration was estimated through three fully connected (FC) layers with 1,024, 512, 256 units followed by a unidirectional LSTM network with 128 memory blocks. 
	Based on this estimated duration, the phoneme-level linguistic features were then upsampled to frame-level adding two numerical vectors of phoneme duration and its relative position. 
	
	In context embedding, the linguistic features were transformed into high-level context vectors. 
	The module here consisted of three convolution layers with a 10×1 kernel and 512 channels per layer, a bi-directional LSTM network with 512 memory blocks, and an FC layer with 512 units. 
	
	We used a Tacotron 2 decoder network to generate the output acoustic features \cite{Shen2018NaturalTS}. 
	First, the previously generated acoustic features were fed into two FC layers with 256 units (i.e. the PreNet), and those features and the vectors from the context embedding module were then passed through two uni-directional LSTM layers with 1,024 memory blocks followed by two projection layers.
	Finally, to improve generation accuracy, five convolution layers with 5$\times$1 kernels and 512 channels per layer were used as a post-processing network (i.e. the PostNet) to add the residual elements of the generated acoustic features.
    
    Before training, the input and output features were normalized to have zero mean and unit variance. 
    The weights were initialized using \textit{Xavier} initialization and \textit{Adam} optimization was used \cite{xavier2010init, diederik2014adam}.
    The learning rate was scheduled to be decayed from 0.001 to 0.0001 via a decaying rate of 0.33 per 100,000 steps.

	\subsubsection{Vocoding model}

    The architecture of the proposed MbG-ExcitNet comprised three convolutional blocks, each with 10 convolution layers with dilations of 1, 2, 4, and so on, up to 512. 
    The numbers of dilated causal convolution channels and 1$\times$1 convolutions in the residual block were both set to 512, and the number of 1$\times$1 convolution channels between the skip connection and the softmax layer was set to 256. 
    
    Before training, the LSFs in the training set were generated by the pre-trained acoustic model and were used to compose the conditional inputs together with auxiliary parameters extracted from the input speech, such as F0, energy, v/uv, SEW, and REW. 
    % eunwoo (5/16): removed footnote
    %\footnote{ 
    %To align LSFs' frame-rate with the input speech signal at the training stage, the ground-truth duration was used for the acoustic model instead of the predicted one.} 
    It is possible to use auxiliary parameters also generated by the pre-trained acoustic model, but we recommend \minjaeedit{to use} ground-truth observations to avoid generating unstable speech segments. 
    % using --> to use
    The conditional inputs were normalized to have zero mean and unit variance and were duplicated from frame to sample to match the length of the input speech signals \cite{tamamori2017speaker}. 
    The corresponding excitation signals were obtained by passing the input speech signals through the LP analysis filters formed by the generated LSFs. 
    % They were then normalized to range between -1.0 and 1.0 and, finally, encoded by 8-bit µ-law compression. 
    \rminjaeedit{They were then normalized to range between -1.0 and 1.0 followed by 8-bit µ-law encoding.}

    The weights were initialized using Xavier initialization and Adam optimization was used. 
    The learning rate was set to 0.0001, and the batch size was set to 30,000 (1.25 sec).

	\subsubsection{TTS system}
	
    In the synthesis step, all of the acoustic feature vectors were predicted by the acoustic model with the given input text.  
    By inputting these features, the MbG-ExcitNet vocoder generated a discrete symbol of the quantized excitation signal, and its dynamic was recovered via $\mu$-law expansion. 
    Finally, the speech signal was reconstructed by applying the LP synthesis filter to the generated excitation signal.

    %%% TABLE: MOS TEST RESULTS
	\begin{table}[!t]   
	\begin{center}         
	\caption{
	TTS \reunwooedit{naturalness} MOS results with 95\% confidence intervals with respect to the different vocoding models: the best MOS scores are in bold.}
	%\vspace*{-3pt}
	\label{table:mos}
	{\small        
	\begin{tabular}{>{}m{.60\linewidth} c}
%	\begin{tabular}{c| m| c}
	\Xhline{2\arrayrulewidth}
	Index~~~~~~~System  &    MOS \\
			\hline \hline
	Test 1~~~~~~WaveNet       & 3.23$\pm$0.11 \\%\hline
	Test 2~~~~~~ExcitNet       & 4.43$\pm$0.08 \\
	Test 3~~~~~~G-WaveNet     & 3.36$\pm$0.11 \\
	Test 4~~~~~~G-ExcitNet     & 3.29$\pm$0.12 \\%\hline
	Test 5~~~~~~MbG-ExcitNet (ours)   & \textbf{4.57$\pm$0.07}  \\
	\hline
	Test 6~~~~~~Raw         & 4.66$\pm$0.07 \\
			\Xhline{2\arrayrulewidth}
	\end{tabular}}          
	\end{center}         
	\vspace*{7pt}
	\end{table}
    %%% TABLE: MOS TEST RESULTS    
    
    \subsection{Evaluations}

    To evaluate the perceptual quality of the proposed system, \reunwooedit{naturalness} MOS tests were performed\footnote{Generated audio samples are available at the following URL:\\ \url{https://sewplay.github.io/demos/mbg_excitnet}} by asking 13 native Korean speakers to make quality judgments about the synthesized speech samples using the following five responses: 1 = Bad; 2 = Poor; 3 = Fair; 4 = Good; and 5 = Excellent. 
    %Note that the listening tests were performed in an acoustically isolated room using a Sennheiser HD650 headphone.
    In total, 20 utterances were randomly selected from the test set and were synthesized using the different generation models. 
    In particular, the speech samples synthesized by the below conventional vocoding methods were evaluated together to confirm performance differences:
    \begin{itemize}
    \item {\bf WaveNet}: Plain WaveNet vocoder \cite{tamamori2017speaker}
    \item {\bf ExcitNet}: Plain ExcitNet vocoder \cite{song2019excitnet}
    \item {\bf G-WaveNet}: WaveNet vocoder trained with generated acoustic parameters \cite{Shen2018NaturalTS}
    \item {\bf G-ExcitNet}: ExcitNet vocoder trained with generated acoustic parameters
    \end{itemize}    
    The G-ExcitNet vocoder was configured similarly to the proposed MbG-ExcitNet, but its target excitation was extracted from the ground-truth spectral parameters.
    
    Table 1 presents the MOS test results for the TTS systems with respect to the different vocoding models, and the analysis can be summarized as follows: First, when training vocoding models using ground-truth acoustic parameters, ExcitNet performed better than WaveNet (Tests 1 and 2). 
    This implies that ExcitNet's adaptive spectral filter is beneficial to reconstruct a more accurate speech signal \cite{song2019excitnet}.
    Second, training the model with generated parameters provided better perceptual quality than using the ground-truth approach in WaveNet (Tests 1 and 3), but vice versa in ExcitNet (Tests 2 and 4). 
    This result confirms that target excitation should be replaced by considering the acoustic model's estimation errors in excitation-based methods. 
    Lastly, the proposed MbG-ExcitNet performed best across the different vocoders (Tests 5 \eunwooedit{and the others}). 
    % test 5 --> tests 5 and the others
    Because the MbG training strategy guided the vocoding model to compensate for errors from the acoustic model, it was possible to significantly improve synthesis accuracy. 
    Consequently, the TTS system with the proposed MbG-ExcitNet vocoder achieved \ttsmos MOS.
    
    To further verify the \minjaeedit{effectiveness} of the proposed method, we designed additional experiments to refine the initialization of MbG-ExcitNet's model weights. 
    % effect --> effectiveness
    Since the MbG training process utilizes generated spectral parameters and the corresponding excitation signals as the input conditions and target outputs, respectively, it may be difficult to capture the speech signals' original characteristics. 
    We therefore adopted a transfer learning method \cite{oquab2014learning} through which the MbG-ExcitNet was initialized by the plain ExcitNet model whose own weights were optimized by ground-truth speech spectra and excitations. 
    All weights were then fine-tuned by the MbG framework. 
    As a result, it was possible to guide the entire training process to learn the characteristics of both the original and the generated speech segments. 
    
\begin{figure}[!t]
    \centerline{\epsfig{figure=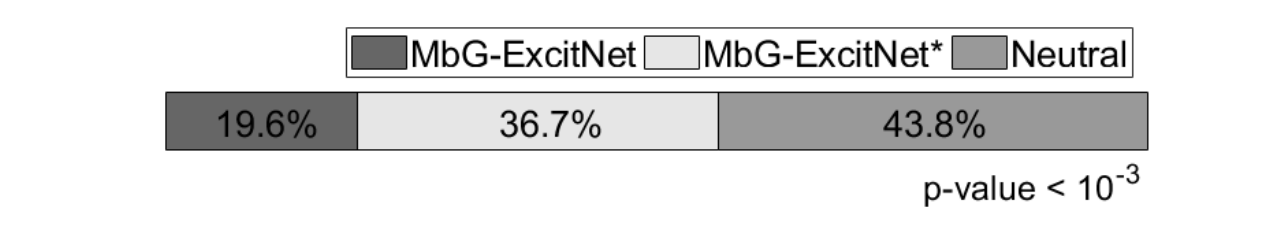,width=90mm}}
%    \vspace*{-6pt}  
    \caption{A/B/X preference comparison of MbG-ExcitNet and its initialization-refined version, MbG-ExcitNet$^{*}$.}
    \label{fig:abx}

    \end{figure}

    Fig.~\ref{fig:abx} depicts the results of an A/B/X preference test between the proposed MbG-ExcitNet and this initialization-refined version (MbG-ExcitNet$^{*}$). 
    The setup for this test was the same as for the MOS assessment except that listeners were asked to rate the quality preference of the synthesized speech samples. 
    The results confirm that the initialization-refined system provided better perceptual quality than the originally proposed MbG-ExcitNet. 
    This \eunwooedit{confirms} that adopting a transfer learning method is advantageous to generating more natural speech signal in an MbG-structured TTS system.
    % implies-->confirms
    
%%%%%%%%%%%%%%%%%%% Fig: abx %%%%%%%%%%%%%%%%%%%%%%%%%%%%%%  
%%%%%%%%%%%%%%%%%%% Fig: abx %%%%%%%%%%%%%%%%%%%%%%%%%%%%%%   
    
\section{Conclusions}
    This paper has proposed a high-quality neural TTS system that incorporates an MbG structure into the ExcitNet vocoder. 
    The MbG-ExcitNet back-end was optimized to learn excitation output distributions while simultaneously compensating for estimation errors from the acoustic model front-end. 
    As such, the proposed method was effective in minimizing the mismatch between the acoustic model and the vocoder. 
    The experimental results verified that a TTS system with the proposed MbG-ExcitNet vocoder performed significantly better than conventional systems with similarly configured WaveNet vocoders. 
    Future research should include extending the framework into speech synthesis systems based on WaveRNN and/or WaveGlow vocoders.

\section{Acknowledgements}
The authors would like to thank \minjaeedit{Hyungseob} Lim, Kyungguen Byun, Seyun Um, and \minjaeedit{Suhyeon Oh} at DSP\&AI Lab., Yonsei University, Seoul, Korea, for their support.
% 이름 확인

\bibliographystyle{IEEEtran}

\bibliography{mybib}

% \begin{thebibliography}{9}
% \bibitem[1]{Davis80-COP}
%   S.\ B.\ Davis and P.\ Mermelstein,
%   ``Comparison of parametric representation for monosyllabic word recognition in continuously spoken sentences,''
%   \textit{IEEE Transactions on Acoustics, Speech and Signal Processing}, vol.~28, no.~4, pp.~357--366, 1980.
% \bibitem[2]{Rabiner89-ATO}
%   L.\ R.\ Rabiner,
%   ``A tutorial on hidden Markov models and selected applications in speech recognition,''
%   \textit{Proceedings of the IEEE}, vol.~77, no.~2, pp.~257-286, 1989.
% \bibitem[3]{Hastie09-TEO}
%   T.\ Hastie, R.\ Tibshirani, and J.\ Friedman,
%   \textit{The Elements of Statistical Learning -- Data Mining, Inference, and Prediction}.
%   New York: Springer, 2009.
% \bibitem[4]{YourName17-XXX}
%   F.\ Lastname1, F.\ Lastname2, and F.\ Lastname3,
%   ``Title of your INTERSPEECH 2020 publication,''
%   in \textit{Interspeech 2020 -- 20\textsuperscript{th} Annual Conference of the International Speech Communication Association, September 15-19, Graz, Austria, Proceedings, Proceedings}, 2020, pp.~100--104.
% \end{thebibliography}

\end{document}